# Phase-selective orbital-charge conversion in $MoTe_2$

J. L. Costa[1], E. Santos[1], G. R. Gallo[2], G. Rodrigues-Junior[2], R. O. Cunha[2], E. L. T. França[1], R. Cardias[3], T. G. Rappoport[3], J. B. S. Mendes[2]*, and A. Azevedo[1]*

[1]*Departamento de Física, Universidade Federal de Pernambuco, 50670-901, Recife, PE, Brazil.*
[2]*Departamento de Física, Universidade Federal de Viçosa, 36570-900, Viçosa, MG, Brazil.*
[3]*Centro Brasileiro de Pesquisas Físicas (CBPF), 22290-180 Rio de Janeiro, RJ, Brazil*

**Abstract**

Two-dimensional transition metal dichalcogenides are promising for spin-orbitronics, yet their orbital transport remains largely unexplored. Here we show that the orbitronic response of $MoTe_2$ is governed by a thickness-driven structural phase transition. Sputtered thin films undergo a crossover near 4.5 nm, stabilizing in metallic 1T′ form below this thickness and semiconducting 2H form above it. Raman and scanning tunneling spectroscopy confirm the structural and electronic transition, revealing gapless ultrathin films and a finite gap in thicker samples. Spin-pumping measurements detect an additional transverse charge-conversion signal only in metallic 1T′-$MoTe_2$, consistent with first-principles calculations identifying a dominant orbital Rashba-Edelstein response.

*E-mail: joaquim.mendes@ufv.br, antonio.azevedo@ufpe.br

Since the groundbreaking isolation of single-layer graphene from graphite [1,2], the field of two-dimensional (2D) materials has expanded rapidly, enabling the discovery of numerous systems with rich physical properties and technological potential. Among them, transition metal dichalcogenides (TMDs) have attracted particular attention due to their versatile electronic structure and strong spin-orbit coupling. Their integration into next-generation nanoelectronic and spintronic devices, however, remains critically dependent on scalable synthesis of uniform, high-quality films. Monolayer TMDs are especially attractive for electronics because their atomic-scale thickness improves electrostatic control and suppresses leakage currents [3-7]. In spintronics, strong spin-orbit coupling and inversion symmetry breaking at monolayer interfaces can generate Rashba effects, spin-momentum locking, and efficient charge-to-spin conversion via the spin Hall effect (SHE), making TMDs promising for spin-orbit torque devices and related technologies [8-14].

TMDs, with general formula $MX_2$, consist of a transition-metal layer sandwiched between two chalcogen layers. Their electronic properties are highly sensitive to structural phase, making phase transitions a central topic of study. $MoTe_2$ is a notable example, exhibiting semiconducting 2H, metallic 1T′, and orthorhombic $T_d$ phases, the latter being type-II Weyl semimetal with topologically protected gapless surface states [15-22]. This polymorphism makes $MoTe_2$ an ideal platform for studying structure-property relations and phase transitions driven by temperature, strain, gating, or growth conditions [23,24]. Previous studies have demonstrated a controllable transition between the 2H and 1T' phases through external perturbations and scalable growth, highlighting phase engineering in $MoTe_2$ as feasible and technologically relevant. Because both phases are intrinsically stable, this introduces the fundamental question of how structural control influences spin and orbital transport.

In this Letter, we investigate the structural, spin, and orbital transport properties of $MoTe_2$ thin films grown by DC magnetron sputtering. Although sputtering is unconventional for two-dimensional materials, it offers clear advantages for wafer-scale fabrication and device integration. We find a thickness-driven structural phase transition: below approximately $4-4.5$ nm, $MoTe_2$ stabilizes in the metallic 1T' phase, whereas thicker films stabilize in the semiconducting 2H order. Raman spectroscopy identifies this crossover, and scanning tunneling spectroscopy reveals the corresponding evolution from gapless to gapped electronic behavior. In ultrathin films, substrate-induced mismatch, interfacial energy, and sputtering-induced strain favor 1T′ stabilization; as thickness increases, these effects relax and the bulk-preferred stable 2H phase emerges. Using spin and orbital pumping, we probe angular-momentum-to-charge conversion across both phases. No measurable spin-to-charge conversion is detected in either the 1T′ or 2H phases of $MoTe_2$. In contrast, an additional transverse charge-conversion signal, attributed to orbital-to-charge conversion appears only in metallic 1T′-$MoTe_2$ within yttrium iron garnet ($Y_3Fe_5O_{12}$, YIG)/Pt/$MoTe_2$ heterostructures. First-principles calculations support these observations, revealing enhanced orbital response in 1T′ and suppressed response in 2H.

Fig. 1(a) shows the grazing-incidence X-ray diffraction (GIXRD) pattern from a 50 nm $MoTe_2$ film deposited on YIG, confirming the polycrystalline growth in the hexagonal 2H phase. The diffraction pattern is consistent with preferential c-axis oriented growth and shows no evidence of competing 1T′ or $T_d$ phases. Ultrathin films (< 4 nm) yield weak XRD signals, making structural identification difficult. Subsequently, we used Raman spectroscopy to probe films of different thicknesses, as this technique is highly effective for distinguishing crystalline phases of TMDs due to their distinct vibrational symmetries [25-27]. Fig. 1(b) shows Raman spectra for YIG/$MoTe_2$ films with thicknesses of 15 nm and 3 nm. The 15 nm film exhibits the in-plane $E_{1g}$ and $E_{2g}^1$ modes near $125\ \mathrm{cm}^{-1}$ and $235\ \mathrm{cm}^{-1}$, characteristic of the semiconducting 2H phase [27]. In contrast, the 3 nm film displays the shear/tilt $B_g$ mode near $160\ \mathrm{cm}^{-1}$, together with the $A_u$ ($\sim 110\ \mathrm{cm}^{-1}$) and $A_g$ ($\sim 128$ and $\sim 260\ \mathrm{cm}^{-1}$) modes, which are characteristic signatures of the metallic 1T′ phase [27]. These results demonstrate a thickness-dependent structural transition, with thicker films (above 4.5 nm) favoring 2H-$MoTe_2$ and ultrathin films (below 4.5 nm) stabilizing 1T′-$MoTe_2$. To prevent oxidation, all samples were capped with a 2 nm Pt layer, whose thickness is justified in the next section. Unlike previous studies, where 2H-1T' transitions were induced by electrostatic doping [28], our results show that sputtered ultrathin films intrinsically stabilize the 1T' phase below a critical thickness, enabling wafer-scale phase engineering without post-growth processing. Substrate-induced strain and interfacial effects likely contribute to this stabilization, although stoichiometry-related effects in ultrathin films cannot be excluded. We next examined the microstructure and interfaces of the full heterostructures by high-resolution cross-sectional scanning transmission electron microscopy (STEM) (Figs. 1(c, d)). The layered morphology confirms the high structural quality of the films and the integrity of the YIG/Pt(2)/$MoTe_2$($t$)/Pt(2) stacks, $t = 3$ nm and 10 nm, respectively. The Pt layers exhibit textured crystalline structure and the YIG/Pt(2)/$MoTe_2$(10)/Pt(2) sample (Fig. 1 (f)) shows lamellar features consistent with well-oriented $MoTe_2$ in the 2H phase. In addition, the YIG layer is monocrystalline with well-oriented [111] growth (see Supplemental Material [29]). Most importantly, STEM reveals atomically sharp and well-defined interfaces throughout the heterostructure, a prerequisite for efficient spin transparency and, even more critically, robust orbital injection.

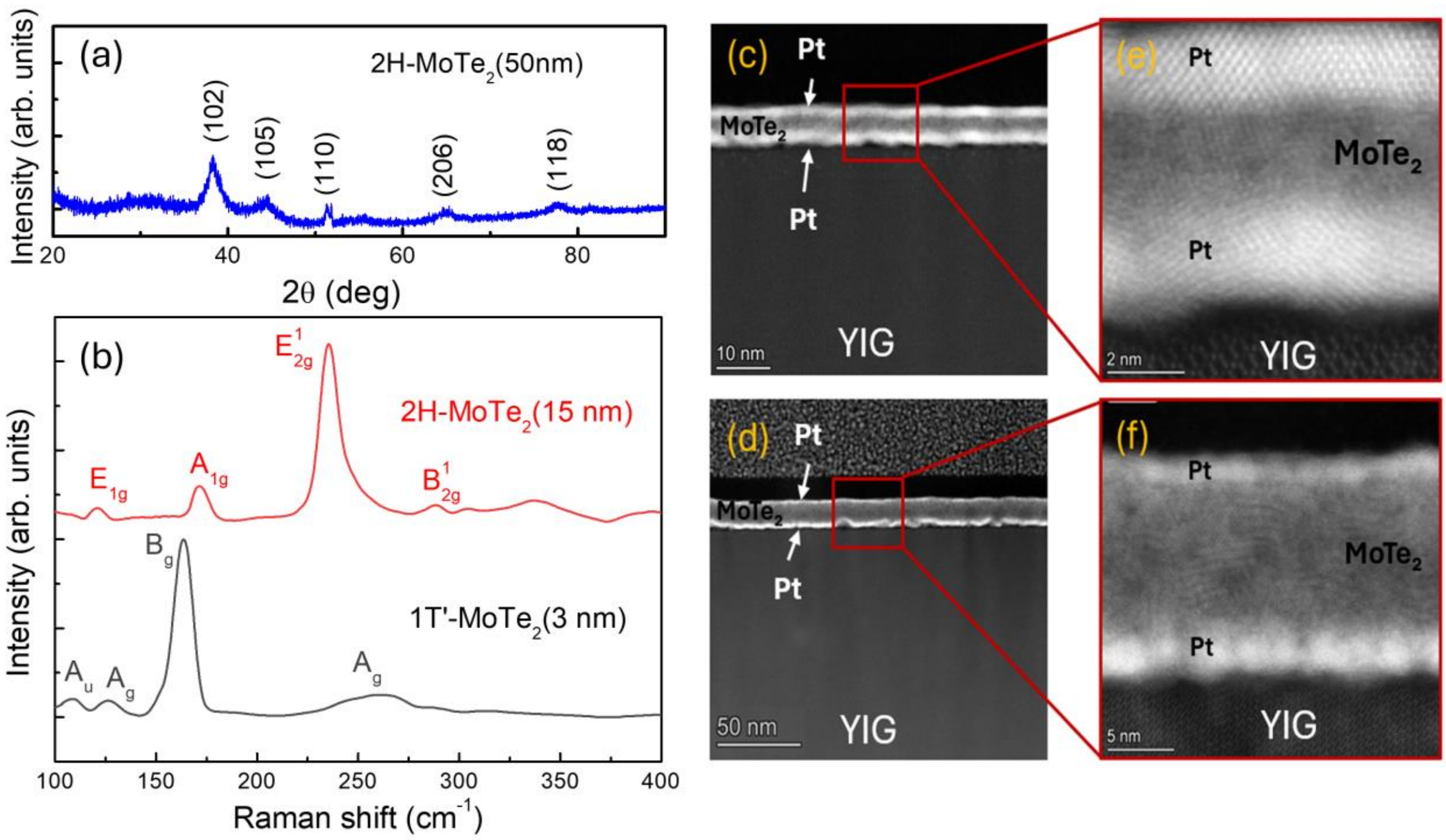

**FIG. 1.** (a) X-ray diffraction pattern of a 50 nm as-deposited $MoTe_2$ film on YIG (111), confirming polycrystalline growth in the hexagonal 2H phase. (b) Raman spectra of $MoTe_2$ films in the 2H (red) and 1T′ (black) crystalline phases. The 15 nm film exhibits in-plane vibrational modes ($E_{1g}, E_{2g}^1$) typical of 2H TMDs, whereas the 3 nm film shows distortion-related and in-plane modes ($B_g, A_u, A_g$) consistent with the 1T′ phase. (c, d) Cross-section high resolution STEM images of YIG/Pt(2)/$MoTe_2$(3)/Pt(2) and YIG/Pt(2)/$MoTe_2$(10)/Pt(2) heterostructures, respectively. (e, f) Enlarged views of the highlighted $MoTe_2$, confirming film continuity and crystalline order.

To further investigate the structural phases of sputtered $MoTe_2$, scanning tunneling microscopy/spectroscopy (STM/STS) was employed to probe the local electronic structure of the films, as displayed in Fig. 2. The STS measurements reveal a pronounced thickness dependence of the electronic properties. For the 10 nm thick film (surface morphology displayed in Fig. 2(a)), the measured dI/dV spectrum presented in Fig. 2(b) shows a pronounced suppression of conductance around the Fermi level, indicative of semiconducting behavior. The energy gap is estimated to be approximately 1.03 eV. This value is consistent with the density of states calculated using Density Functional Theory (DFT), which also exhibits a clear gap around $E_F$, characteristic of semiconducting 2H-$MoTe_2$. In contrast, the 3 nm $MoTe_2$ film (surface morphology displayed in Fig. 2(d)) displays a strongly suppressed but finite differential conductance near the Fermi level, without any well-defined energy gap, indicating a nearly gapless semimetallic electronic structure characteristic of the 1T′ phase, as can be seen in Fig. 2(e). The minimum in dI/dV is slightly shifted from zero bias, which may arise from charge transfer, residual doping, or tip-sample work-function differences. This interpretation is supported by the DFT-calculated DOS (Fig. 2(f)), which shows a pronounced minimum around $E_F$, yet no finite gap, in sharp contrast to the gapped DOS observed for the 2H phase (Fig. 2(c)). These experimental and theoretical results consistently demonstrate a thickness-driven phase transition in sputtered $MoTe_2$, from semiconducting 2H-$MoTe_2$ in thicker films (10 nm) to nearly gapless semimetallic 1T′ phase in ultrathin films (3 nm), in agreement with previous reports of controllable phase transitions in the material [30]. A more comprehensive analysis of the STM/STS results is provided in the Supplemental Material [29].

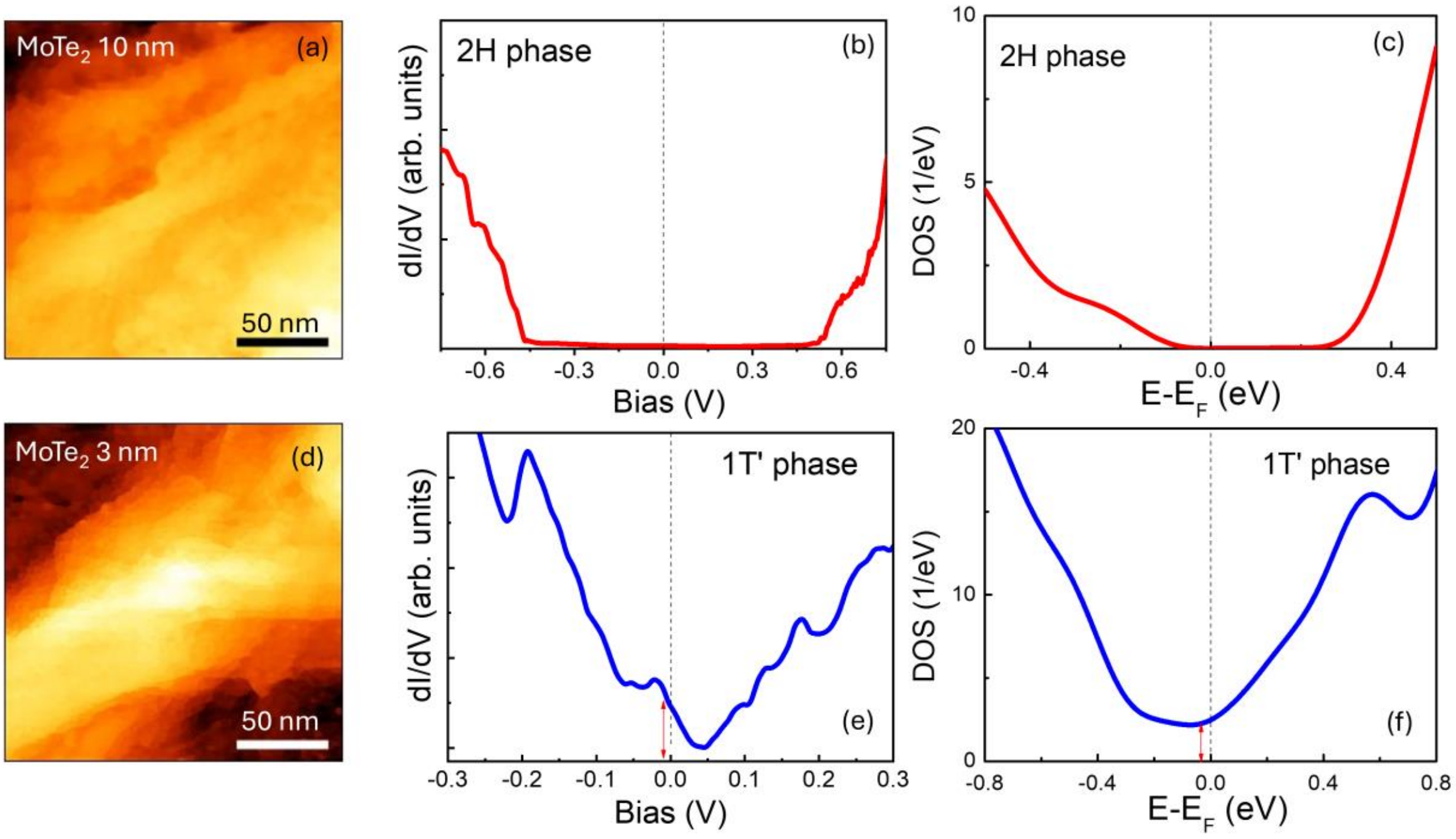

**FIG. 2.** (a, b) STM surface topography and corresponding local density of states of a 10 nm-thick $MoTe_2$ film, probed by STS through differential conductance (dI/dV) measurements. A well-defined bang gap of approximately $E_g = 1.03\ eV$ is observed, indicating semiconducting behavior consistent with the 2H crystalline phase. (c) Density of states calculated by DFT for 2H-phase $MoTe_2$, showing a clear gap around the Fermi level, in good agreement with the experimental STS result. (d, e) STM surface topography and the corresponding dI/dV spectrum of 3 nm-thick $MoTe_2$ film, exhibiting a strongly suppressed but finite conductance near the Fermi level, indicative of a nearly gapless semimetallic electronic structure characteristic of the 1T′ phase. (f) DFT-calculated density of states for 1T′-$MoTe_2$, showing a pronounced minimum but finite DOS at $E_F$, consistent with the experimental STS observation.

To investigate the spin and orbital transport properties in $MoTe_2$, we employed the spin-pumping technique driven by ferromagnetic resonance (SP-FMR), with specific experimental details provided in the Supplemental Material. In this technique, the precessing magnetization of a ferromagnetic layer (FM) injects a spin current ($\vec{J}_S$) into an adjacent normal metal (NM). In the presence of SOC, this spin current is converted into a transverse charge current ($\vec{J}_c$) due to the inverse spin Hall effect (ISHE) [31-34], mathematically described by the equation $\vec{J}_c^{ISHE} = (2e/\hbar)\theta_{SH}(\vec{J}_S \times \hat{\sigma}_S)$ in the bulk of the NM, or the inverse spin Rashba-Edelstein effect (ISREE), described by the equation $\vec{J}_C^{ISREE} = \lambda_{ISREE}(\hat{z} \times \delta\vec{S})$ at the interface, where $\hat{\sigma}_S$ is the spin polarization, $\delta\vec{S}$ represents the non-equilibrium spin density at the interface and $\hat{z}$ is the unit vector perpendicular to the surface of the film, and $\theta_{SH}$ and $\lambda_{ISREE}$ quantify the efficiencies of bulk and interfacial spin-charge conversion, respectively. [35,36].

In addition to ISHE and ISREE, which convert spin angular momentum flux into a transverse charge current, it is also possible to generate an orbital angular momentum flux, or an orbital current ($\vec{J}_L$), and detect a transverse charge current. Similarly to the spin effects, the conversion can be driven either at the interface or the bulk. These effects are the inverse orbital Hall effect (IOHE) and the inverse orbital Rashba Edelstein effect (IOREE) [36], which are described by the equations $\vec{J}_c^{IOHE} = (2e/\hbar)\theta_{OH}(\vec{J}_L \times \hat{\sigma}_L)$ for IOHE [36-41], and $\vec{J}_C^{IOREE} = \lambda_{IOREE}(\hat{z} \times \delta\vec{L})$ for IOREE [36,41,42], where $\delta\vec{L}$ represents the interface orbital accumulation, and $\theta_{OH}$ and $\lambda_{IOREE}$ quantify the efficiencies of bulk and interfacial orbital-charge conversion, respectively. The orbital polarization ($\hat{\sigma}_L$), in some cases, can be determined by the $\hat{\sigma}_S$ in the material, due to the interaction mediated by the SOC. Several experimental studies report an efficient method for generating orbital current from the YIG/Pt bilayer [36-41]. In summary, the method is based on the precession of the YIG magnetization, which induces spin accumulation at the YIG/Pt interface. This accumulation diffuses into the Pt as a spin current. Owing to the strong SOC in Pt, both spin and orbital currents are generated and diffuse through the YIG/Pt/NM structure toward the NM layer, where they may be converted into a charge current through reciprocal spin or orbital phenomena.

We designed the heterostructures shown in Fig. 3(a) to probe the spin and orbital conversion mechanisms in $MoTe_2$. A thin layer of Pt (2) deposited on YIG plays a dual role: first, it converts part of the spin current injected from YIG under FMR into a charge current via the ISHE; second, owing to its strong spin-orbit coupling, it enables the generation of spin-orbital intertwined currents that propagate toward the $MoTe_2$ layer. As a result, the transverse voltage measured in YIG/Pt/$MoTe_2$ heterostructures contains the ISHE contribution from Pt together with an additional conversion contribution associated with $MoTe_2$. Fig. 3(b) shows representative SP-FMR signals for three main control samples: YIG/Pt(2) (orange and black symbols), YIG/Pt(2)/$MoTe_2$(3)/Pt(2) (blue and red symbols), and YIG/$MoTe_2$(3)/Pt(2) (inset). By comparing the signals with and without the $MoTe_2$ layer, the contribution of the $MoTe_2$ orbital effect can be isolated. While the YIG/Pt bilayer exhibits the expected ISHE signal, the YIG/Pt(2)/$MoTe_2$(3)/Pt(2) heterostructure exhibits SP-FMR signal with amplitude approximately twice as large as those of the YIG/Pt reference sample. ISHE signals generated in $MoTe_2$ is absent, as shown in the inset of Fig. 3(b). The top Pt(2) layer acts as a covering layer to prevent $MoTe_2$ oxidation, and positive signals are observed at $\phi = 0°$ and negative signals at $\phi = 180°$, where the reversal of the signal polarity is due to the reversal of the spin (or orbital) polarization. Surprisingly, the insertion of a 3 nm $MoTe_2$ layer between YIG and the protective Pt layer (YIG/$MoTe_2$/Pt) yields a negligible SP-FMR signal (bottom left inset in Fig. 3(b)), indicating that $MoTe_2$ does not allow the spin flux to reach the Pt layer, as evidenced by the small signal observed in this sample. The SP-FMR signals are expressed as electric current, dividing the measured voltage by the electrical resistance between the electrodes, as illustrated in Fig. 3(a). The lower right inset in Fig. 3(b) shows the FMR absorption curve of the YIG(400 nm) film, with a resonance field around 2.6 kOe.

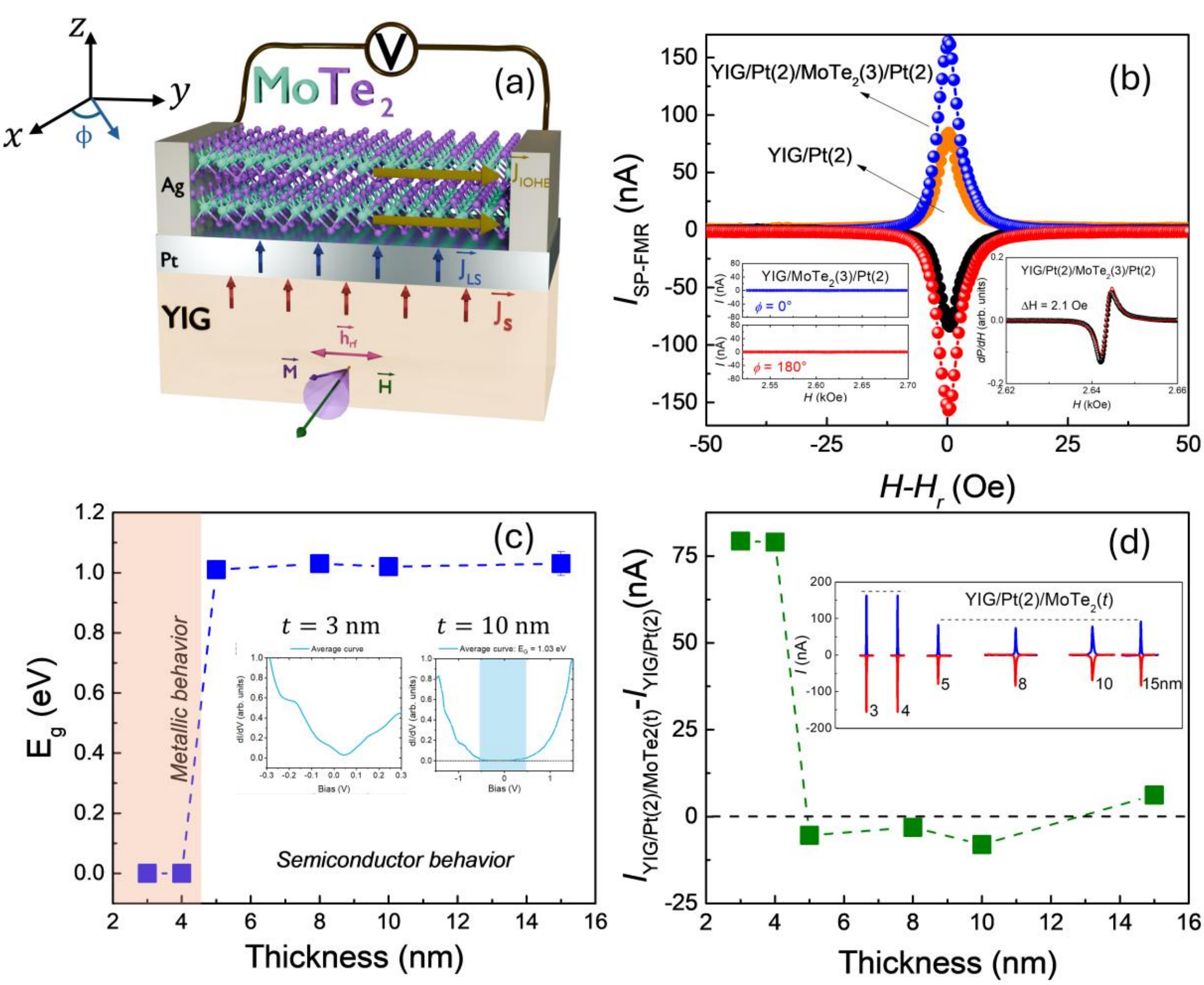

**FIG. 3.** (a) Schematic of the YIG/Pt(2)/$MoTe_2$ heterostructure and the spin/orbital-to-charge conversion mechanisms. (b) SP-FMR signals measured for YIG/Pt(2) and YIG/Pt(2)/$MoTe_2$(3)/Pt(2) at $\phi = 0°$ and $180°$. Insets: SP-FMR signal of YIG/$MoTe_2$(3)/Pt(2), showing negligible ISHE, and the corresponding FMR absorption spectrum. (c) Band gap energy Eg extracted from STS as a function of $MoTe_2$ thickness, showing metallic behavior for $t = 3$ nm, 4 nm and semiconducting behavior for thicker films. (d) Difference between the SP-FMR charge currents measured in YIG/Pt(2)/$MoTe_2$(t)/Pt(2) and YIG/Pt(2), isolating the $MoTe_2$ contribution. Orbital-to-charge conversion is observed only for 3 nm and 4 nm films.

To directly correlate the transport behavior to the electronic structure of the materials, we performed STS on the full set of YIG/$MoTe_2$($t_{MoTe_2}$) samples. Since the differential conductance $dI/dV$ is proportional to the local density of states, it provides direct information on the material band gap. For semiconductors, $dI/dV$ exhibits a zero-conductance region around zero bias, whereas metals show finite conductance. The insets of Fig. 3(c) show representative $dI/dV$ curves for $t_{\mathrm{MoTe_2}} = 3$ nm and 10 nm. The thinner sample exhibits metallic behavior, while the thicker one shows semiconducting behavior with a band gap of approximately 1.03 eV. The thickness-dependent evolution of the bang gap (blue symbols in Fig. 3(c)), reveals a structural phase transition in sputtered $MoTe_2$, consistent with a crossover between the 1T′ and 2H phases previously identified by Raman spectroscopy (Fig. 1(b)). Specifically, $MoTe_2$ is metallic for $t_{\mathrm{MoTe_2}} < 5$ nm and semiconducting for $t_{\mathrm{MoTe_2}} \geq 5$ nm. The full STS data is provided in the Supplemental Material [29]. The central result is summarized in Fig. 3(d), which plots the difference in the SP-FMR charge current between YIG/Pt(2)/$MoTe_2$($t$)/Pt(2) and the reference sample YIG/Pt(2). As demonstrated by the YIG/$MoTe_2$(3)/Pt(2) control sample, $MoTe_2$ alone exhibits negligible spin-to-charge conversion under our experimental conditions. Therefore, the measured difference reflects an additional charge-conversion contribution associated with the $MoTe_2$ layer. A clear signal enhancement is observed only for $MoTe_2$ < 5 nm, indicating that this additional contribution emerges exclusively in the metallic 1T' regime. Remarkably, the enhanced signal is detected only for $MoTe_2$ thicknesses of 3 and 4 nm, while it becomes negligible for thicker films in the semiconducting 2H phase. Finally, the comparable amplitudes observed for the 3 nm and 4 nm samples, without a clear thickness dependence, are consistent with a predominantly interfacial orbital-to-charge conversion mechanism in 1T' $MoTe_2$, as expected from IOREE. The absence of signal enhancement for $MoTe_2$ ≥ 5 nm correlates directly with the sizable band gap of the 2H phase, approximately 1.03 eV, which suppresses the density of itinerant carriers at the Fermi level and strongly reduces charge-conversion processes. In contrast, the metallic character of the 1T' phase enables an efficient conversion channel in the ultrathin regime. These results show that the additional charge-conversion response in sputtered $MoTe_2$ is not universal, but highly sensitive to its structural and electronic phase. The strong correlation between gapless electronic states and the observation of a finite additional signal indicates that charge conversion in $MoTe_2$ is intimately linked to its electronic structure. Motivated by this connection, we calculate the current-induced spin and orbital accumulations in semi-infinite $MoTe_2$ slabs within a first-principles linear-response framework based on the Kubo-Bastin formalism. Fig. 4 shows the energy dependence of the spin and orbital Rashba-Edelstein responses for semi-infinite $MoTe_2$ in the 2H and 1T' phases. In addition to the total response, Fig. 4 presents the layer-resolved distribution, showing that the induced spin and orbital polarizations are strongly concentrated at the outermost layer. This behavior is consistent with an interfacial charge-conversion mechanism and makes the Rashba-Edelstein response the most relevant quantity for the experimental heterostructures studied here. Since the experimental films are polycrystalline and no measurable dependence on in-plane orientation is observed, the theoretical results for the 1T' phase are presented as an average over the two considered in-plane orientations. Although the calculation does not explicitly include the Pt/$MoTe_2$ interface, it captures the intrinsic phase dependence of $MoTe_2$ and therefore provides microscopic insight into the relative charge-conversion efficiency expected at the interface

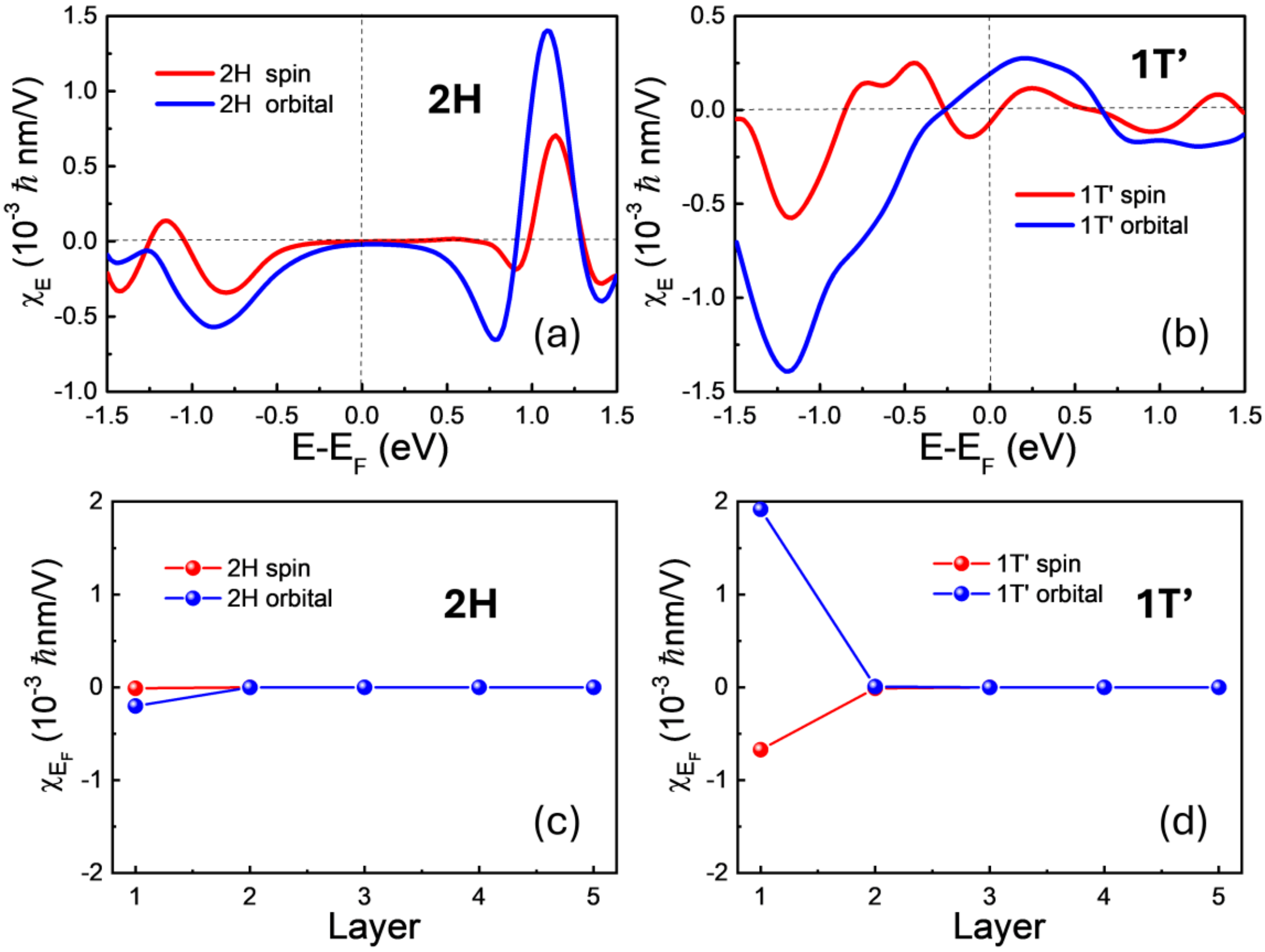

**FIG. 4.** Spin and orbital Rashba-Edelstein response of semi-infinite $MoTe_2$ in the 2H and 1T' phases. Panels (a) and (b) display the energy dependence of the orbital (blue) and spin (red) Rashba-Edelstein susceptibilities for the 2H and 1T' phases, respectively. While both responses are strongly suppressed near the Fermi level in the semiconducting 2H phase, the metallic 1T' phase exhibits a pronounced response, dominated by the orbital contribution. Panels (c) and (d) show the corresponding layer-resolved response at the Fermi level. In both phases, the induced response is concentrated mainly at the outermost layer, indicating its interfacial character. For 1T'-$MoTe_2$, the theoretical values correspond to the average over the two in-plane orientations considered in the calculations.

A key result is that, in both structural phases, the interfacial spin and orbital accumulations associated with the Rashba-Edelstein effect are substantially larger than the corresponding interfacial accumulations originating from the spin and orbital Hall effects. For this reason, the main text focuses on the Rashba-Edelstein response as the dominant mechanism for charge conversion in the present heterostructures. The two phases of $MoTe_2$, however, behave very differently near the Fermi level. In the 2H phase, both the spin and orbital Rashba-Edelstein responses are strongly suppressed in the vicinity of the Fermi level, consistent with its semiconducting character. In contrast, the 1T' phase exhibits a sizable response near the Fermi level, reflecting its metallic electronic structure. Importantly, in this phase the orbital contribution is much larger than the spin contribution over the energy range most relevant to the experiment.

These results provide a natural interpretation of the experimental findings. Experimentally, an additional charge-conversion signal is observed only for ultrathin $MoTe_2$ films in the metallic 1T' phase, whereas thicker samples in the semiconducting 2H phase do not show a comparable enhancement. The strong suppression of both spin and orbital Rashba-Edelstein responses in 2H-$MoTe_2$ is consistent with the absence of an additional signal in this phase. In contrast, the sizable response obtained for 1T'-$MoTe_2$ supports the emergence of an additional conversion channel in the ultrathin regime.

We have demonstrated that the orbital charge-conversion in sputtered $MoTe_2$ thin films can be selectively activated by controlling their structural phase through film thickness. By employing a suite of characterization techniques, we have established a clear thickness-dependent phase transition: below approximately 4.5 nm, $MoTe_2$ stabilizes in a metallic 1T′ structure, whereas thicker films adopt the semiconducting 2H phase. Using SP-FMR technique, we observed a pronounced inverse orbital Rashba-Edelstein effect only in the 1T′ phase, with no significant inverse spin Hall effect or Rashba-Edelstein detected in either phase. These results are consistently supported by first-principles calculations, which reveal a substantial Rashba-Edelstein effect near the Fermi level in the 1T′ phase, in stark contrast to its suppression in the gapped 2H phase. Our study reveals that orbital transport in $MoTe_2$ is intrinsically linked to its electronic structure and crystallographic phase, with the metallic 1T′ phase hosting efficient orbital-to-charge conversion. This phase-dependent behavior highlights $MoTe_2$ as a tunable material platform for orbitronics, where orbital currents can be controlled via structural engineering. Moreover, the scalability of the sputtering technique enables wafer-compatible phase stabilization without external stimuli, which is advantageous for device integration. These results not only establish $MoTe_2$ as a promising candidate for phase-engineered orbital-to-charge conversion devices but also suggest broader implications for the design of orbital-based spintronic components using two-dimensional materials with structural polymorphism. Future work may explore dynamic phase switching via strain, doping, or gating to achieve real-time control of orbital transport in functional spin-orbitronic circuits.

**Acknowledgments**

This research is supported by Coordenação de Aperfeiçoamento de Pessoal de Nível Superior (CAPES), Conselho Nacional de Desenvolvimento Científico e Tecnológico (CNPq), Financiadora de Estudos e Projetos (FINEP), Fundação de Amparo à Pesquisa do Estado de Minas Gerais (FAPEMIG) - Rede de Pesquisa em Materiais 2D and Rede de Nanomagnetismo, Fundação de Amparo à Ciência e Tecnologia do Estado de Pernambuco (FACEPE), Universidade Federal de Pernambuco, Multiuser Laboratory Facilities of DF-UFPE, INCT em Nanocarbono e Materiais 2D, INCT of Spintronics and Advanced Magnetic Nanostructures (INCT-SpinNanoMag), CNPq 406836/2022-1. This research used the facilities of the Brazilian Nanotechnology National Laboratory (LNNano), part of the Brazilian Centre for Research in Energy and Materials (CNPEM). Therefore, the authors acknowledge LNNano/CNPEM for advanced infrastructure and technical support (Proposal No. 20243516, TEM-FEG facility).

**Data availability statement**

The data that support the findings of this study are available from the corresponding author upon reasonable request.

**Notes**

The authors declare no competing financial interest.

---